%% file: gasmotions.tex
\documentclass[12pt]{article}
\usepackage{times}
\usepackage{geometry}
\usepackage{floatflt}
\usepackage{wrapfig}
\usepackage[utf8]{inputenc}
\usepackage{enumitem,amssymb}
\usepackage{ragged2e}
\newlist{thematic}{itemize}{8}
\setlist[thematic]{label=$\square$}
\usepackage{pifont}
\newcommand{\cmark}{\ding{51}}%
\newcommand{\done}{\rlap{$\square$}{\raisebox{2pt}{\large\hspace{1pt}\cmark}}%
\hspace{-2.5pt}}

\input{macros.tex}

\usepackage{jheppub}   
\usepackage[dvipsnames,svgnames,x11names]{xcolor}
\usepackage{sectsty}
\usepackage[innercaption]{sidecap}
\sidecaptionvpos{figure}{c}

\newcommand\xmm{{\sl XMM-Newton}}

\newcommand\axis{{\sl AXIS}}
\newcommand\athena{{\sl Athena}}
\newcommand\lynx{{\sl Lynx}}

\begin{document}
\raggedright
\huge
Astro2020 Science White Paper \linebreak

Probing Macro-Scale Gas Motions and Turbulence in Diffuse Cosmic Plasmas \linebreak
\normalsize

\noindent \textbf{Thematic Areas:} \hspace*{60pt} $\square$ Planetary Systems \hspace*{10pt} $\square$ Star and Planet Formation \hspace*{20pt}\linebreak
$\square$ Formation and Evolution of Compact Objects \hspace*{31pt} $\done$  Cosmology and Fundamental Physics \linebreak
  $\square$  Stars and Stellar Evolution \hspace*{1pt} $\square$ Resolved Stellar Populations and their Environments \hspace*{40pt} \linebreak
 $\done$    Galaxy Evolution   \hspace*{45pt} $\square$             Multi-Messenger Astronomy and Astrophysics \hspace*{65pt} \linebreak
  
\textbf{Principal Author:}

Name: Esra Bulbul 
 \linebreak						
Institution:  Center for Astrophysics $|$ Harvard \& Smithsonian
 \linebreak
Email: ebulbul@cfa.harvard.edu
 \linebreak
Phone:  +1 617-496-7523
 \linebreak

\textbf{Co-authors:} 
  \linebreak
 Massimo~Gaspari (Princeton Univ.), 
 Gabriella~Alvarez (CfA),
 Camille~Avestruz (Univ. of Chicago),
 Mark~Bautz (MIT),
 Brad~Benson (Univ. of Chicago),
 Veronica~Biffi (CfA),
 Douglas~Burke (CfA),
 Urmila~Chadayammuri (CfA), 
 Eugene~Churazov (MPA),
 Nicolas~Clerc (IRAP/Toulouse),
 Edoardo~Cucchetti (IRAP/Toulouse), 
 Dominique~Eckert (Univ. of Geneva),
 Stefano~Ettori (INAF-OAS Bologna),
 Bill~Forman (CfA),
 Fabio~Gastaldello (INAF-IASF Milano), 
 Vittorio~Ghirardini (CfA), 
 Ralph~Kraft (CfA), 
 Maxim~Markevitch (NASA/GSFC),
 Mike~McDonald (MIT),  
 Eric~Miller (MIT), 
 Tony~Mroczkowski (ESO), 
 Daisuke~Nagai (Yale Univ.), 
 Paul~Nulsen (CfA), 
 Gabriel~W.~Pratt (IRFU/CEA), 
 Scott~Randall (CfA),
 Thomas~Reiprich (Univ. of Bonn),
 Mauro~Roncarelli (INAF/ Bologna), 
 Aurora~Simionescu (SRON), 
 Randall~Smith (CfA),
 Grant~Tremblay (CfA),
 Stephen~Walker (NASA/GSFC),
 Irina~Zhuravleva (Univ. of Chicago),
 John~ZuHone (CfA)

 \justify
\textbf{Abstract:}

Clusters of galaxies, the largest collapsed structures in the Universe, are located at the intersection of extended filaments of baryons and dark matter. Cosmological accretion onto clusters through large scale filaments adds material at cluster outskirts. Kinetic energy in the form of bulk motions and turbulence due to this accretion provides a form of pressure support against gravity, supplemental to thermal pressure. Significant amount of non-thermal pressure support could bias cluster masses derived assuming hydrostatic equilibrium, the primary proxy for cluster cosmology studies. Sensitive measurements of Doppler broadening and shift of astrophysical lines, and the relative fluctuations in thermodynamical quantities (e.g., density, pressure, and entropy) are primary diagnostic tools. Forthcoming planned and proposed X-ray (with large etendue, throughput, and high spectral resolution) and SZ observatories will provide crucial information on the assembly and virialisation processes of clusters, involving turbulent eddies cascading at various spatial scales and larger gas bulk motions in their external regions to the depth or their potential wells.

\pagebreak
\clearpage

\pagenumbering{arabic}
     
\section{Introduction}
\vspace{-1mm}
Clusters of galaxies, the largest collapsed structures in the Universe, are located at the intersection of extended filaments of baryons and dark matter. Cosmological accretion onto clusters through large scale filaments adds material at cluster outskirts, where infalling gas forms a shock \citep[for recent reviews, see][]{reiprich13,walker19}. However, due to this filamentary structure of the cosmic web, the gas does not accrete uniformly, and generates inhomogeneities and gas clumps \citep{nagai2011,vazza13,lau15} that are observed in deep X-ray exposures of cluster outskirts \citep{morandi14,ich15,bulbul16,sim17}. 
Many substructures are incorporated into the cluster material via ram-pressure and dynamical stripping, while some structures are dense and resilient enough to penetrate to deep radii, such as massive galaxies and small groups \citep{zinger16}. The gas motions due to this infall depend sensitively on the physical processes operating in the intracluster medium (ICM) on such scales \citep{rasia14}, such as the cluster magnetic field, viscosity, and thermal conduction.

\begin{SCfigure}[1.5][h!]
  \hspace{2mm}
  \includegraphics[totalheight=2.9in]{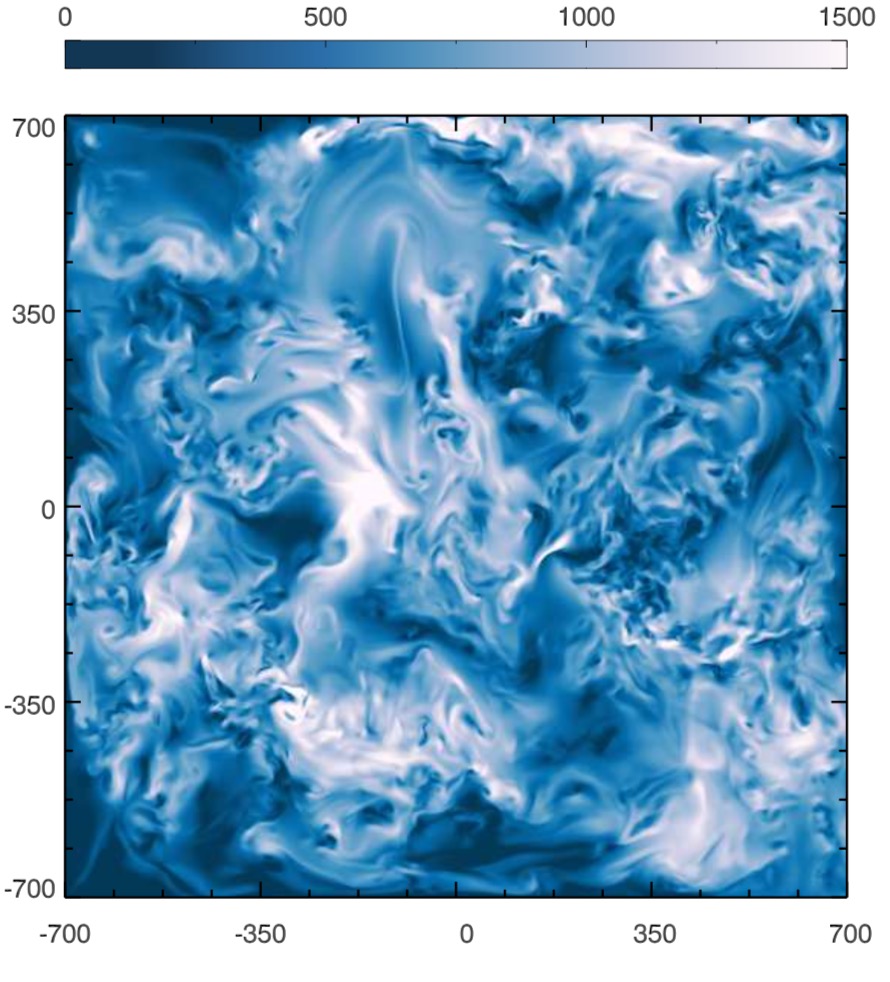}
   \caption{Simulation of the turbulent intracluster medium in a massive Coma-like cluster. The cross-section shown is the total velocity magnitude in the units of km/s  \citep[adapted from][]{gaspari14}. A significant amount of turbulence and gas motions could bias hydrostatic cluster masses, a primary proxy used in cluster cosmology studies. Forthcoming planned and proposed X-ray and SZ observatories will provide crucial information on the assembly and virialization processes in the outskirts of clusters. 
   }\label{fig:fig0}
\end{SCfigure}
\vspace{-4mm}

Kinetic energy in the form of bulk motions and turbulence provides a form of pressure support against gravity, supplemental to thermal pressure.
Cosmological simulations indicate that subsonic chaotic motions are ubiquitous, with turbulent pressure support in the range of 5-35 percent of the total ICM pressure from relaxed to merging clusters \citep[e.g.,][]{nelson14,shi15,vazza18}.  Significant amount of non-thermal pressure support could bias cluster masses derived assuming hydrostatic equilibrium, the primary proxy for cluster cosmology studies, by up to 20\% \citep{lau09,biffi16}. {\bf Understanding the details of hydrodynamical processes in the low surface brightness outskirts of the macro-scale structures would significantly advance our understanding of how structure forms and evolves in the Universe as well as the use of galaxy clusters as cosmological probes.}

In this perspective, the combined use of X-ray data and observations
of the Sunyaev-Zeldovich (SZ) effect represent an optimal way to explore
the physics of the ICM and estimate the impact of gas motions on
hydrostatic mass estimates, especially in the outermost regions where
most of the cluster mass resides.
 Sensitive measurements of Doppler broadening, shifts in astrophysical lines, and the relative fluctuations in thermodynamical quantities (e.g., density, pressure, and entropy) are primary diagnostic tools. These constraints are limited to the cluster cores at modest energy resolution (E/$\Delta$E$\sim$50) with current generation of X-ray observatories. Measurements of these quantities out to the splashback radius can only be made with X-ray telescopes with much larger (10$\times$) effective area and higher energy-dispersive spectral resolution (E/$\Delta$E$>$500) \citep[e.g.,][]{mansfield17}. The measurements of hydro/thermodynamical properties with future high spatial and spectral resolution X-ray missions (\athena, \lynx, and \axis) will yield a more complete view of ICM dynamics and non-thermal pressure support. 

Complementary to X-rays, the thermal Sunyaev-Zeldovich effect provides unique capabilities for probing astrophysical processes at high redshifts and out to the low-density regions in the outskirts of galaxy clusters \citep[for a recent review, see][]{Mroczkowski2019}. In the following sections, we highlight how combining X-ray observations with data from current and future SZ instruments (e.g.  SPT-3G, AdvACT, the Simons Observatory, CMB-S4, MUSTANG2, NIKA2, TolTEC, ALMA) will provide a major leap forward in understanding the underlying physical processes in cluster outskirts. Additionally, the observed synchrotron emission in cluster radio halos and radio relics can be explained by stochastic acceleration generated by turbulent motions in the ICM and particle acceleration at shocks found in cluster outskirts \citep[see][for a review]{vanWeeren19}. Measurements of the thermodynamics and kinematics in the X-ray emitting ICM provide complementary information to the surveys in the radio band (e.g., LOFAR and SKA), that is key in allowing us to answer fundamental questions regarding the physics of particle acceleration in diffuse cosmic plasmas \citep{pfrommer08}.
      
\section{Fluctuation Power Spectrum}

\noindent Spatially-resolved measurements of the relative fluctuations in X-ray and tSZ 2D distributions of surface brightness, pressure, and entropy are primary probes of the amplitudes of the velocity field \citep{Churazov.ea:12,gaspari13, gaspari14, Zhuravleva.ea:15, khatri16}. The large-scale bulk
motions resulting from mergers and accretion have high Reynolds
numbers, causing them to drive chaotic turbulent motions \citep{shi18}. This implies the formation of large-scale eddies that transport the kinetic energy to smaller scales via progressively smaller vortices. The `inertial range' of the turbulence cascade is finally broken at the dissipation scale ($< 1$ kpc), where viscosity is significant enough to convert kinetic into thermal energy. 

\begin{figure}[ht!]
\hspace{-5mm}
\includegraphics[totalheight=3in]{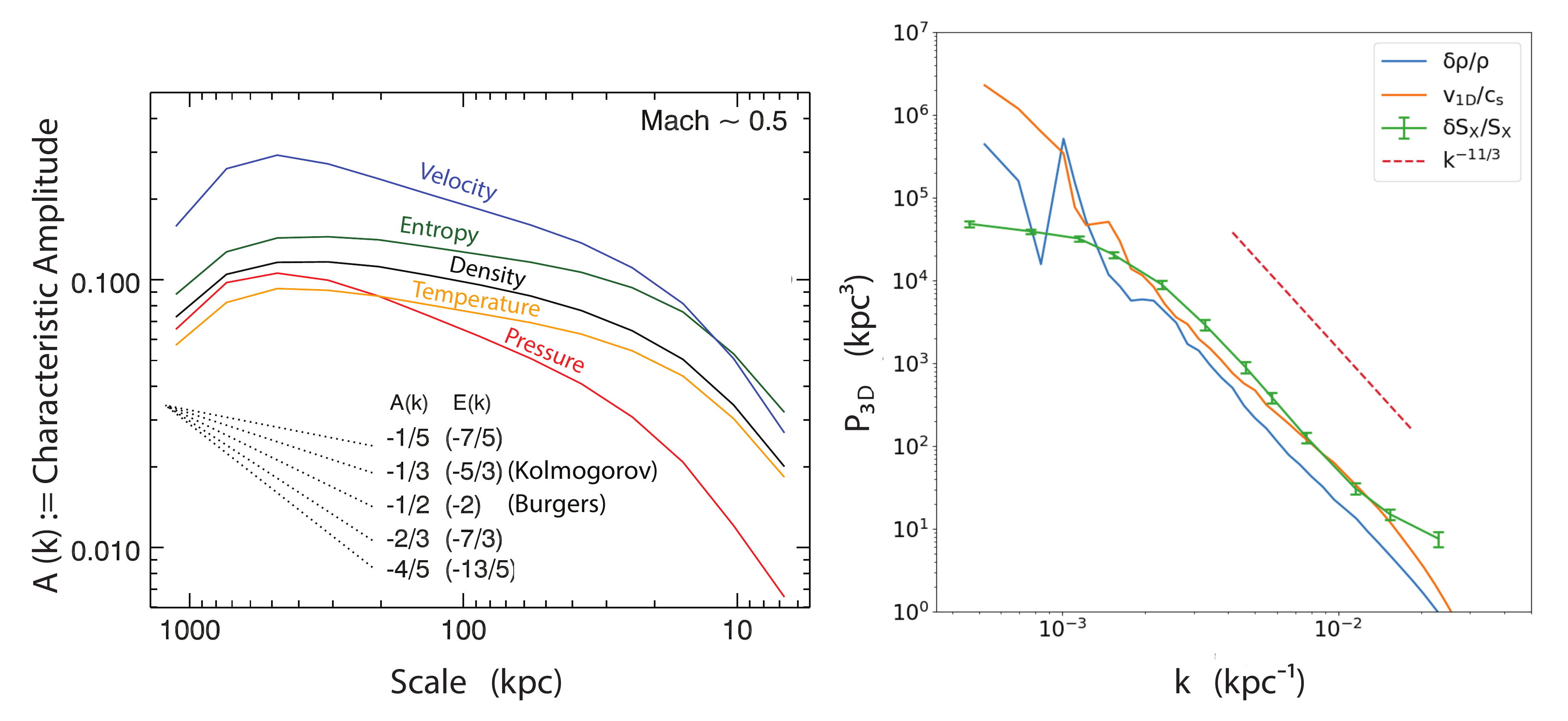}
\caption{Left Panel: Power spectra of thermodynamic and velocity fluctuations (entropy, density, pressure, temperature), which arise from a turbulent ICM in the pure hydrodynamical case with a merger injection scale of $L\sim$\,500\,kpc \citep{gaspari14}. Right Panel: Density power spectrum obtained from 100~ks \athena\ WFI observations of a simulated galaxy cluster at z=0.1 is shown in green. The velocity and density power spectra obtained from the hydrodynamical simulations of the same cluster are shown in blue and orange. The 2D (green) curve has been normalized by the integral of the power spectrum of the emission measure. Accurate measurements of perturbations in density and pressure are vital tools for recovering the physics of the faint outskirts of clusters. Density fluctuations in the outskirts ($\sim$1~Mpc) of a nearby cluster will be recovered with a precision of 4\% in a relatively short \athena\ WFI observations. The power-law slope (Kolmogorov slope shown in dashed line) will be measured from the slope of the power spectrum.}
\label{fig:p2d}
\end{figure}

The formation and evolution of the turbulence cascade is best unveiled in Fourier space, where each wave mode ($k\propto 1/l$) can be separated. Figure~\ref{fig:p2d} (left) shows the Fourier power spectrum (PS) of turbulent velocities and all the relative thermodynamic perturbations (e.g. $\delta \rho/\rho$ for gas density; black line), arising from an intermediate subsonic turbulence with 3D ${\rm Mach} \equiv \sigma_v/c_{\rm s}\sim 0.5$. Evidently, the PS of all thermodynamic perturbations correlate to some degree with the velocity power spectrum. In a stratified halo such as the ICM, the turbulent Mach number linearly increases with the amplitude of density and surface brightness fluctuations \citep{gaspari13}. 
The power spectrum of gas motions can be recovered from the surface brightness (SB) and pressure fluctuations seen in X-ray and tSZ measurements in the cluster outskirts. The slope of the PS is tightly related to the transport processes in the ICM \citep{gaspari14}. 
In the pure hydrodynamical regime, the PS slope of density fluctuations follows a Kolmogorov ($E(k) \propto k^{-5/3}$) or slightly shallower cascade. However, in the presence of significant thermal conduction, the density and temperature fluctuations at small scales are washed out, and the related spectral slope steepens toward the Burgers-like case ($E(k) \propto k^{-2}$). Thereby, measuring the PS slope is a key diagnostic to assess the level of conductivity in the medium.

Applications of the PS method find a substantially suppressed conduction compared to the Spitzer value \citep{gaspari13,eckert14,eckert17b,DeGrandi16}. The Fourier PS method also reveals the thermodynamic mode of the ICM fluctuations. Combining the fluctuation constraints on density, pressure, and entropy provides crucial information on whether the underlying thermo-hydrodynamical processes follow an isobaric or adiabatic effective equation of state \citep{zhuravleva18}. Specifically, the major modes in the ICM can be grouped into buoyancy waves versus sound waves: the former are tied mostly to entropy fluctuations, while the latter are associated with major pressure fluctuations \citep{gaspari14}. The transition between the two regimes typically occurs at Mach$ >$ 0.5. Furthermore, significant clumping in cluster outskirts may lead to an increase in the surface brightness power spectrum on relevant scales, as already suggested (although with a low significance) from \textit{Chandra} observations of the Perseus Cluster \citep{urban14}. Cross-correlating the density and velocity power spectra will reveal the properties of the surviving clumps which fall onto clusters from surrounding filaments.

Initial studies on PS of X-ray surface brightness fluctuations remain limited to the inner cluster cores \citep{schuecker04,Churazov.ea:12,gaspari13}.
The macro-scale PS analysis of pressure fluctuations through low-noise \textit{Planck} tSZ observations up to the virial radius of Coma cluster revealed for the first time significant amount of non-thermal pressure support in cluster outskirts up to $\sim$45\% \citep{khatri16}. On the other hand, a larger sample of (more relaxed) \xmm\ clusters indicates a more modest average level of non-thermal pressure support \citep{eckert17,Siegel2018}. Limitations of current telescopes due to their small mirror effective and collecting area, restrict the studies of density fluctuations in X-ray faint cluster outskirts. Accurate measurements of small scale perturbations (down to dissipation scales of $\sim$5~kpc) require telescopes with high angular resolution, large etendue (or throughput, $\propto A \Omega$), and low instrumental background. The Wide Field Imager (WFI) on board \athena\ will allow the first accurate measurements of  perturbations in thermodynamical properties down to 7~kpc scales in the local Universe \citep{nandra13,Meidinger14}. Figure~\ref{fig:p2d} (right) demonstrates that the density fluctuations will be recovered with an accuracy of 4\% at $\sim$1~Mpc in a relatively short 100~ks \athena\ WFI synthetic observation of a nearby Coma-like cluster. The high quality statistics provided by the WFI will allow us to probe the smallest accessible spatial scales, pushing the constraints toward the Coulomb mean free path around which hydrodynamical turbulence should dissipate. The higher spatial resolution High Definition X-ray Imager on board of \lynx\  and \axis\ will extend fluctuations measurements of the level of fluctuations on much smaller scales (~1 kpc) to be determined \citep{gaskin15}. Thermal SZ studies will also elucidate the properties of the ICM in the outskirts by inferring the pressure fluctuations and testing the pressure equilibrium of clumps, and measuring, and even imaging, accretion shocks. Additionally, kinetic SZ (kSZ) measurements may provide constraints on turbulent motions within the ICM \citep[as shown in][]{Sayers2019}. For many intermediate and high-z systems, such deep, resolved tSZ and kSZ measurements may only be enabled by a new, large-aperture ($>30$ meter), wide-field ($>1^\circ$) mm/submm-wave telescopes such as  \href{http://atlast-telescope.org}{AtLAST} \cite{Bertoldi2018,deBreuck2018,Hargrave2018,Klaassen2018,Mroczkowski2018,Otarola2019}. Such a large facility with a wide field of view is required for sensitive measurements of the tSZ and kSZ that probe out to, and beyond, the splashback radius of a cluster \citep{mroczkowski19b}.

\section{Direct X-ray Spectroscopic Measurements}

Dynamical processes impact the X-ray spectrum of the hot ICM, and more specifically the line profiles of the emission from heavy elements (e.g., iron). 
 Lines are broadened by turbulent motions in the gas, and shifted by bulk motions which, when integrated over the line of sight,
may produce detectable structure in line profiles. The characterization of emission lines from the hot gas in clusters is therefore a direct measurement of the ICM velocity field. The centroid shift in the energies of emission lines probe the bulk velocity of the gas  along the line of sight \cite[e.g.,][]{nagai13}, and will constrain the large-scale velocity field and the  amplitude of the turbulent power spectrum at the injection scale \citep[e.g.,][] {zuhone16}. The imprint of turbulence and small scale motions is contained of the line shape. 

The choice of an ``unperturbed'' model with respect to which the surface brightness fluctuation power spectra are computed affects the inferred amplitude of the perturbations and is therefore a source of significant systematic uncertainty. 
It is therefore challenging to determine the exact strength of the perturbations, particularly on large scales. Direct measurements through high-resolution spectroscopy are of vital importance to validate and quantify the related uncertainties in the power spectrum analyses. Once these limitations are understood, the results from surface brightness fluctuations analysis and amplitude of the turbulent power spectrum can be combined into a powerful test of the interesting underlying physics in the ICM. 

\begin{figure}
  \begin{minipage}[c]{0.47\textwidth}
    \includegraphics[width=\textwidth]{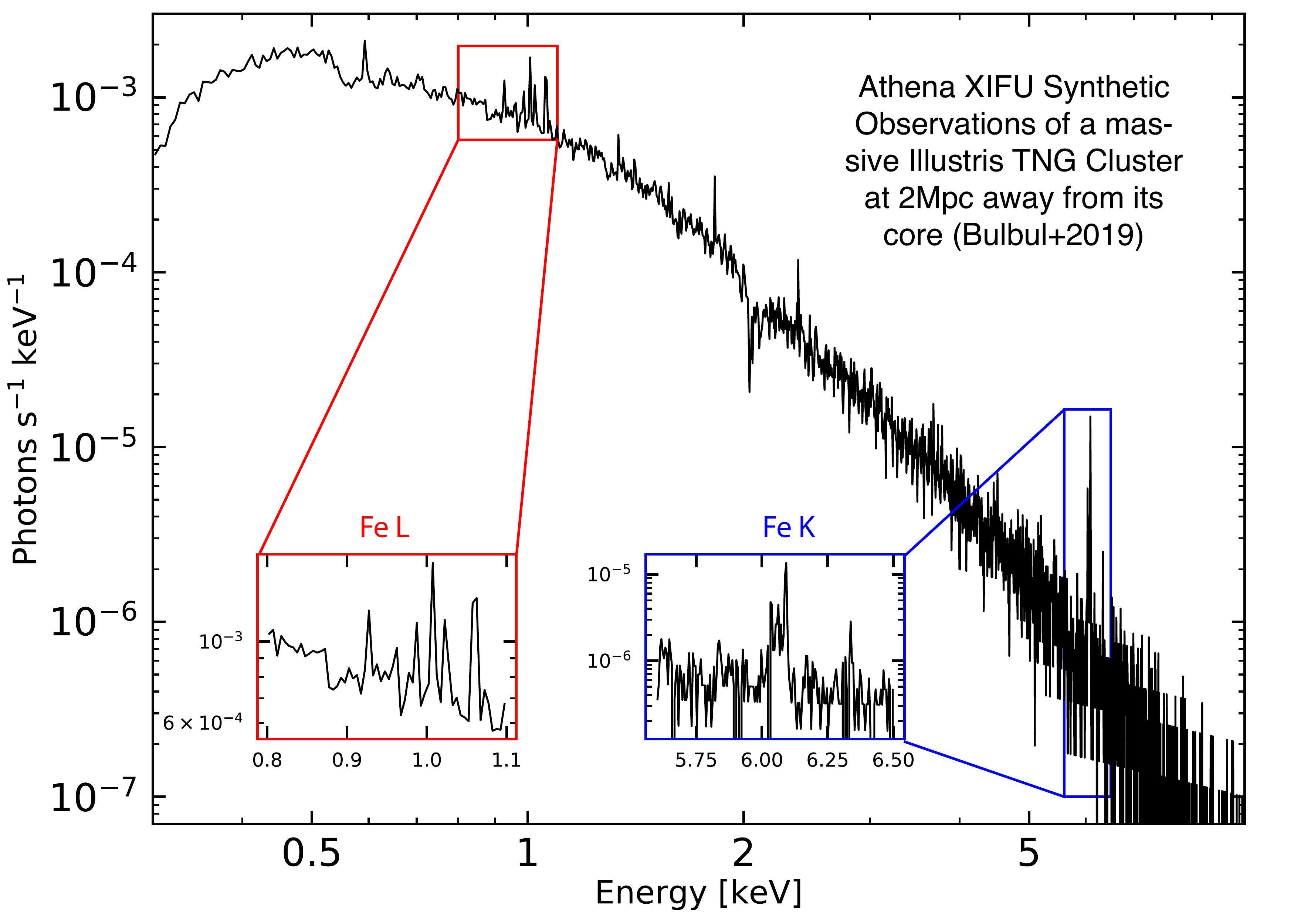}
  \end{minipage}\hfill
  \begin{minipage}[c]{0.45\textwidth}
    \caption{1~Ms synthetic \athena\ X-IFU observations at $R_{200}$ (2~Mpc away from the core) of a massive cluster ($M_{200}=9\times10^{14}\ M_{\odot}$) at a redshift of 0.07 adopted from the Illustris TNG  simulations \citep{nelson18}. The insets are zooms over the iron L and K complexes.
    } \label{fig:xifu}
  \end{minipage}
  \vspace{-2mm}
\end{figure}

The energy resolution of current CCDs and present-generation grating spectrometers is on the of order 1000 km s$^{-1}$, limiting the ability to measure ICM motion spectroscopically.
Previous attempts of constraining turbulence have been limited to cluster cores through energy-dispersive spectroscopy \citep{bulbul2012, Sanders.ea:13a, Pinto.ea:15, Bambic.ea:18,hitomi16, Hitomi18_turbulence,ZuHone2018}. 
The {\it Hitomi} X-ray observatory \citep{takahashi14} (with spectral resolution $\sim$5~eV) measured bulk and turbulent gas motions for the first time, through shifting and broadening of the 6.7 keV Fe XXV K$\alpha$ line in the core of the Perseus cluster \citep{hitomi18}, providing physical insights into the nature of turbulent gas motions driven by cosmic weather and AGN feedback \citep{lau17,bourne17}. The forthcoming \textit{XRISM}/Resolve instrument (E/$\Delta$E$\sim$1600) scheduled to be launched in 2022 will perform measurements of Doppler broadening of emission lines on limited spatial scales ($\sim$100 kpc bins)  to intermediate cluster radii \citep{tashiro18, Guainazzi2018} and measure the hydrostatic mass bias due to the non-thermal pressure  (out to $R_{500}$) of nearby clusters \citep{ota18, simionescu19}. However, it will be unable to extend these measurements to cluster outskirts. Accurate measurements of line morphology in faint cluster outskirts require deep observations with higher effective area (10$\times$, EA$>$1 m$^{2}$) and higher energy-dispersive spectral resolution (E/$\Delta$E$>$500) instruments. The X-ray Integral Field Unit, X-IFU \citep{barret16, barret18} onboard  \athena\ is designed to provide spatially-resolved observations of the sky with unprecedented joint spectral and spatial resolution. In conjunction with the 1.4\,m$^2$ effective area (at 1\,keV) of the \athena's mirror, it will provide a transformational leap forward for direct measurements of turbulent and bulk motion velocities. The spectral resolution of 2.5\,eV FWHM and the precision on the energy scale of 0.4\,eV will ensure precise measurements down to 10-20 km/s velocities at $1 \sigma$ confidence level (see Figure \ref{fig:xifu}). \athena\ XIFU and the microcalorimeter on board \lynx\ will provide a comprehensive understanding of the ICM velocity power spectrum down to several kpc scales in the nearby Universe and characterize of the kinematics of the ICM at dissipation scales \citep[see][for feasibility studies for {\it Athena}]{ettori13, roncarelli18, cucchetti18}.

\vspace{-2mm}
\section{Concluding Remarks}
\vspace{-1mm}
Combining fluctuations in X-ray and SZ images and high resolution spectroscopy provide a unique method for measuring the thermodynamic and kinematic properties of the ICM in the low-density regions in the outskirts of galaxy clusters. Cross correlating power spectrum analysis with direct measurements of turbulent cascade via Doppler broadening of emission lines through high resolution spectroscopy is the most powerful approach in determining the micro-physics (e.g., conduction and viscosity) of the ICM.
Forthcoming planned and proposed X-ray observatories with large etendue and high spectral resolution will have the ability to recover the power spectra of the velocity fields for a large range of spatial scales. The joint observations will provide crucial information on the assembly and virialization processes of massive halos, involving turbulent eddies cascading at various spatial scales and larger gas bulk motions contributing to the thermalization of astrophysical objects from their external regions to the depth of their potential wells. These observations will also aid precision cosmology through the measurements of hydrostatic mass bias, one of the major systematic uncertainties in cluster-based cosmological constraints. 

\newpage
\bibliographystyle{unsrturltrunc8}
\bibliography{literature}

\end{document}

%% file: macros.tex
\newcommand{\Moon}{$M_{Moon}$}                          

\newcommand{\apjl}{\rm ApJL}
\newcommand{\apjs}{\rm ApJS}

\newcommand{\aap}{\rm A$\&$A}

\def\ao{{\it Appl.~Opt.}}           

\def\3he{$^3{\rm He}$}
\def\arcdeg{\hbox{$^\circ$}}

\def\lsim{\mathrel{\lower2.5pt\vbox{\lineskip=0pt\baselineskip=0pt
           \hbox{$<$}\hbox{$\sim$}}}}

\def\gsim{\mathrel{\lower2.5pt\vbox{\lineskip=0pt\baselineskip=0pt
           \hbox{$>$}\hbox{$\sim$}}}}

%% file: gasmotions.bbl
\begin{thebibliography}{10}

\bibitem{reiprich13}
T.~H. {Reiprich}, K.~{Basu}, S.~{Ettori}, H.~{Israel}, L.~{Lovisari},
  S.~{Molendi}, E.~{Pointecouteau}, and M.~{Roncarelli}.
\newblock {Outskirts of Galaxy Clusters}.
\newblock {\em \ssr}, 177:195--245, August 2013.
\newblock \href {http://arxiv.org/abs/1303.3286} {\path{arXiv:1303.3286}},
  \href {http://dx.doi.org/10.1007/s11214-013-9983-8}
  {\path{doi:10.1007/s11214-013-9983-8}}.

\bibitem{walker19}
S.~{Walker}, A.~{Simionescu}, D.~{Nagai}, N.~{Okabe}, D.~{Eckert},
  T.~{Mroczkowski}, H.~{Akamatsu}, S.~{Ettori}, et~al.
\newblock {The Physics of Galaxy Cluster Outskirts}.
\newblock {\em \ssr}, 215:7, January 2019.
\newblock \href {http://arxiv.org/abs/1810.00890} {\path{arXiv:1810.00890}},
  \href {http://dx.doi.org/10.1007/s11214-018-0572-8}
  {\path{doi:10.1007/s11214-018-0572-8}}.

\bibitem{nagai2011}
D.~{Nagai} and E.~T. {Lau}.
\newblock {Gas Clumping in the Outskirts of {$\Lambda$}CDM Clusters}.
\newblock {\em \apjl}, 731:L10, April 2011.
\newblock \href {http://arxiv.org/abs/1103.0280} {\path{arXiv:1103.0280}},
  \href {http://dx.doi.org/10.1088/2041-8205/731/1/L10}
  {\path{doi:10.1088/2041-8205/731/1/L10}}.

\bibitem{vazza13}
F.~{Vazza}, D.~{Eckert}, A.~{Simionescu}, M.~{Br{\"u}ggen}, and S.~{Ettori}.
\newblock {Properties of gas clumps and gas clumping factor in the
  intra-cluster medium}.
\newblock {\em \mnras}, 429:799--814, February 2013.
\newblock \href {http://arxiv.org/abs/1211.1695} {\path{arXiv:1211.1695}},
  \href {http://dx.doi.org/10.1093/mnras/sts375}
  {\path{doi:10.1093/mnras/sts375}}.

\bibitem{lau15}
E.~T. {Lau}, D.~{Nagai}, C.~{Avestruz}, K.~{Nelson}, and A.~{Vikhlinin}.
\newblock {Mass Accretion and its Effects on the Self-similarity of Gas
  Profiles in the Outskirts of Galaxy Clusters}.
\newblock {\em \apj}, 806:68, June 2015.
\newblock \href {http://arxiv.org/abs/1411.5361} {\path{arXiv:1411.5361}},
  \href {http://dx.doi.org/10.1088/0004-637X/806/1/68}
  {\path{doi:10.1088/0004-637X/806/1/68}}.

\bibitem{morandi14}
A.~{Morandi} and W.~{Cui}.
\newblock {Measuring the gas clumping in Abell 133}.
\newblock {\em \mnras}, 437:1909--1917, January 2014.
\newblock \href {http://arxiv.org/abs/1306.6336} {\path{arXiv:1306.6336}},
  \href {http://dx.doi.org/10.1093/mnras/stt2021}
  {\path{doi:10.1093/mnras/stt2021}}.

\bibitem{ich15}
Y.~{Ichinohe}, N.~{Werner}, A.~{Simionescu}, S.~W. {Allen}, R.~E.~A. {Canning},
  S.~{Ehlert}, F.~{Mernier}, and T.~{Takahashi}.
\newblock {The growth of the galaxy cluster Abell 85: mergers, shocks,
  stripping and seeding of clumping}.
\newblock {\em \mnras}, 448:2971--2986, April 2015.
\newblock \href {http://arxiv.org/abs/1410.1955} {\path{arXiv:1410.1955}},
  \href {http://dx.doi.org/10.1093/mnras/stv217}
  {\path{doi:10.1093/mnras/stv217}}.

\bibitem{bulbul16}
E.~{Bulbul}, S.~W. {Randall}, M.~{Bayliss}, E.~{Miller}, F.~{Andrade-Santos},
  R.~{Johnson}, M.~{Bautz}, E.~L. {Blanton}, et~al.
\newblock {Probing the Outskirts of the Early-Stage Galaxy Cluster Merger
  A1750}.
\newblock {\em \apj}, 818:131, February 2016.
\newblock \href {http://arxiv.org/abs/1510.00017} {\path{arXiv:1510.00017}},
  \href {http://dx.doi.org/10.3847/0004-637X/818/2/131}
  {\path{doi:10.3847/0004-637X/818/2/131}}.

\bibitem{sim17}
A.~{Simionescu}, N.~{Werner}, A.~{Mantz}, S.~W. {Allen}, and O.~{Urban}.
\newblock {Witnessing the growth of the nearest galaxy cluster: thermodynamics
  of the Virgo Cluster outskirts}.
\newblock {\em \mnras}, 469:1476--1495, August 2017.
\newblock \href {http://arxiv.org/abs/1704.01236} {\path{arXiv:1704.01236}},
  \href {http://dx.doi.org/10.1093/mnras/stx919}
  {\path{doi:10.1093/mnras/stx919}}.

\bibitem{zinger16}
E.~{Zinger}, A.~{Dekel}, Y.~{Birnboim}, A.~{Kravtsov}, and D.~{Nagai}.
\newblock {The role of penetrating gas streams in setting the dynamical state
  of galaxy clusters}.
\newblock {\em \mnras}, 461:412--432, September 2016.
\newblock \href {http://arxiv.org/abs/1510.05388} {\path{arXiv:1510.05388}},
  \href {http://dx.doi.org/10.1093/mnras/stw1283}
  {\path{doi:10.1093/mnras/stw1283}}.

\bibitem{rasia14}
E.~{Rasia}, E.~T. {Lau}, S.~{Borgani}, D.~{Nagai}, K.~{Dolag}, C.~{Avestruz},
  G.~L. {Granato}, P.~{Mazzotta}, et~al.
\newblock {Temperature Structure of the Intracluster Medium from
  Smoothed-particle Hydrodynamics and Adaptive-mesh Refinement Simulations}.
\newblock {\em \apj}, 791:96, August 2014.
\newblock \href {http://arxiv.org/abs/1406.4410} {\path{arXiv:1406.4410}},
  \href {http://dx.doi.org/10.1088/0004-637X/791/2/96}
  {\path{doi:10.1088/0004-637X/791/2/96}}.

\bibitem{gaspari14}
M.~{Gaspari}, E.~{Churazov}, D.~{Nagai}, E.~T. {Lau}, and I.~{Zhuravleva}.
\newblock {The relation between gas density and velocity power spectra in
  galaxy clusters: High-resolution hydrodynamic simulations and the role of
  conduction}.
\newblock {\em \aap}, 569:A67, September 2014.
\newblock \href {http://arxiv.org/abs/1404.5302} {\path{arXiv:1404.5302}},
  \href {http://dx.doi.org/10.1051/0004-6361/201424043}
  {\path{doi:10.1051/0004-6361/201424043}}.

\bibitem{nelson14}
K.~{Nelson}, E.~T. {Lau}, and D.~{Nagai}.
\newblock {Hydrodynamic Simulation of Non-thermal Pressure Profiles of Galaxy
  Clusters}.
\newblock {\em \apj}, 792:25, September 2014.
\newblock \href {http://arxiv.org/abs/1404.4636} {\path{arXiv:1404.4636}},
  \href {http://dx.doi.org/10.1088/0004-637X/792/1/25}
  {\path{doi:10.1088/0004-637X/792/1/25}}.

\bibitem{shi15}
X.~{Shi}, E.~{Komatsu}, K.~{Nelson}, and D.~{Nagai}.
\newblock {Analytical model for non-thermal pressure in galaxy clusters - II.
  Comparison with cosmological hydrodynamics simulation}.
\newblock {\em \mnras}, 448:1020--1029, March 2015.
\newblock \href {http://arxiv.org/abs/1408.3832} {\path{arXiv:1408.3832}},
  \href {http://dx.doi.org/10.1093/mnras/stv036}
  {\path{doi:10.1093/mnras/stv036}}.

\bibitem{vazza18}
F.~{Vazza}, M.~{Angelinelli}, T.~W. {Jones}, D.~{Eckert}, M.~{Br{\"u}ggen},
  G.~{Brunetti}, and C.~{Gheller}.
\newblock {The turbulent pressure support in galaxy clusters revisited}.
\newblock {\em \mnras}, 481:L120--L124, November 2018.
\newblock \href {http://arxiv.org/abs/1809.02690} {\path{arXiv:1809.02690}},
  \href {http://dx.doi.org/10.1093/mnrasl/sly172}
  {\path{doi:10.1093/mnrasl/sly172}}.

\bibitem{lau09}
E.~T. {Lau}, A.~V. {Kravtsov}, and D.~{Nagai}.
\newblock {Residual Gas Motions in the Intracluster Medium and Bias in
  Hydrostatic Measurements of Mass Profiles of Clusters}.
\newblock {\em \apj}, 705:1129--1138, November 2009.
\newblock \href {http://arxiv.org/abs/0903.4895} {\path{arXiv:0903.4895}},
  \href {http://dx.doi.org/10.1088/0004-637X/705/2/1129}
  {\path{doi:10.1088/0004-637X/705/2/1129}}.

\bibitem{biffi16}
V.~{Biffi}, S.~{Borgani}, G.~{Murante}, E.~{Rasia}, S.~{Planelles}, G.~L.
  {Granato}, C.~{Ragone-Figueroa}, A.~M. {Beck}, et~al.
\newblock {On the Nature of Hydrostatic Equilibrium in Galaxy Clusters}.
\newblock {\em \apj}, 827:112, August 2016.
\newblock \href {http://arxiv.org/abs/1606.02293} {\path{arXiv:1606.02293}},
  \href {http://dx.doi.org/10.3847/0004-637X/827/2/112}
  {\path{doi:10.3847/0004-637X/827/2/112}}.

\bibitem{mansfield17}
P.~{Mansfield}, A.~V. {Kravtsov}, and B.~{Diemer}.
\newblock {Splashback Shells of Cold Dark Matter Halos}.
\newblock {\em \apj}, 841:34, May 2017.
\newblock \href {http://arxiv.org/abs/1612.01531} {\path{arXiv:1612.01531}},
  \href {http://dx.doi.org/10.3847/1538-4357/aa7047}
  {\path{doi:10.3847/1538-4357/aa7047}}.

\bibitem{Mroczkowski2019}
T.~{Mroczkowski}, D.~{Nagai}, K.~{Basu}, J.~{Chluba}, J.~{Sayers}, R.~{Adam},
  E.~{Churazov}, A.~{Crites}, et~al.
\newblock {Astrophysics with the Spatially and Spectrally Resolved
  Sunyaev-Zeldovich Effects. A Millimetre/Submillimetre Probe of the Warm and
  Hot Universe}.
\newblock {\em \ssr}, 215:17, February 2019.
\newblock \href {http://arxiv.org/abs/1811.02310} {\path{arXiv:1811.02310}},
  \href {http://dx.doi.org/10.1007/s11214-019-0581-2}
  {\path{doi:10.1007/s11214-019-0581-2}}.

\bibitem{vanWeeren19}
R.~J. {van Weeren}, F.~{de Gasperin}, H.~{Akamatsu}, M.~{Br{\"u}ggen},
  L.~{Feretti}, H.~{Kang}, A.~{Stroe}, and F.~{Zandanel}.
\newblock {Diffuse Radio Emission from Galaxy Clusters}.
\newblock {\em \ssr}, 215:16, February 2019.
\newblock \href {http://arxiv.org/abs/1901.04496} {\path{arXiv:1901.04496}},
  \href {http://dx.doi.org/10.1007/s11214-019-0584-z}
  {\path{doi:10.1007/s11214-019-0584-z}}.

\bibitem{pfrommer08}
C.~{Pfrommer}.
\newblock {Simulating cosmic rays in clusters of galaxies - III. Non-thermal
  scaling relations and comparison to observations}.
\newblock {\em \mnras}, 385:1242--1256, April 2008.
\newblock \href {http://arxiv.org/abs/0707.1693} {\path{arXiv:0707.1693}},
  \href {http://dx.doi.org/10.1111/j.1365-2966.2008.12957.x}
  {\path{doi:10.1111/j.1365-2966.2008.12957.x}}.

\bibitem{Churazov.ea:12}
E.~{Churazov}, A.~{Vikhlinin}, I.~{Zhuravleva}, A.~{Schekochihin},
  I.~{Parrish}, R.~{Sunyaev}, W.~{Forman}, H.~{B{\"o}hringer}, et~al.
\newblock {X-ray surface brightness and gas density fluctuations in the Coma
  cluster}.
\newblock {\em \mnras}, 421:1123--1135, April 2012.
\newblock \href {http://arxiv.org/abs/1110.5875} {\path{arXiv:1110.5875}},
  \href {http://dx.doi.org/10.1111/j.1365-2966.2011.20372.x}
  {\path{doi:10.1111/j.1365-2966.2011.20372.x}}.

\bibitem{gaspari13}
M.~{Gaspari} and E.~{Churazov}.
\newblock {Constraining turbulence and conduction in the hot ICM through
  density perturbations}.
\newblock {\em \aap}, 559:A78, November 2013.
\newblock \href {http://arxiv.org/abs/1307.4397} {\path{arXiv:1307.4397}},
  \href {http://dx.doi.org/10.1051/0004-6361/201322295}
  {\path{doi:10.1051/0004-6361/201322295}}.

\bibitem{Zhuravleva.ea:15}
I.~{Zhuravleva}, E.~{Churazov}, P.~{Ar{\'e}valo}, A.~A. {Schekochihin}, S.~W.
  {Allen}, A.~C. {Fabian}, W.~R. {Forman}, J.~S. {Sanders}, et~al.
\newblock {Gas density fluctuations in the Perseus Cluster: clumping factor and
  velocity power spectrum}.
\newblock {\em \mnras}, 450:4184--4197, July 2015.
\newblock \href {http://arxiv.org/abs/1501.07271} {\path{arXiv:1501.07271}},
  \href {http://dx.doi.org/10.1093/mnras/stv900}
  {\path{doi:10.1093/mnras/stv900}}.

\bibitem{khatri16}
R.~{Khatri} and M.~{Gaspari}.
\newblock {Thermal SZ fluctuations in the ICM: probing turbulence and
  thermodynamics in Coma cluster with Planck}.
\newblock {\em \mnras}, 463:655--669, November 2016.
\newblock \href {http://arxiv.org/abs/1604.03106} {\path{arXiv:1604.03106}},
  \href {http://dx.doi.org/10.1093/mnras/stw2027}
  {\path{doi:10.1093/mnras/stw2027}}.

\bibitem{shi18}
X.~{Shi}, D.~{Nagai}, and E.~T. {Lau}.
\newblock {Multiscale analysis of turbulence evolution in the
  density-stratified intracluster medium}.
\newblock {\em \mnras}, 481:1075--1082, November 2018.
\newblock \href {http://arxiv.org/abs/1806.05056} {\path{arXiv:1806.05056}},
  \href {http://dx.doi.org/10.1093/mnras/sty2340}
  {\path{doi:10.1093/mnras/sty2340}}.

\bibitem{eckert14}
D.~{Eckert}, S.~{Molendi}, M.~{Owers}, M.~{Gaspari}, T.~{Venturi},
  L.~{Rudnick}, S.~{Ettori}, S.~{Paltani}, et~al.
\newblock {The stripping of a galaxy group diving into the massive cluster
  A2142}.
\newblock {\em \aap}, 570:A119, October 2014.
\newblock \href {http://arxiv.org/abs/1408.1394} {\path{arXiv:1408.1394}},
  \href {http://dx.doi.org/10.1051/0004-6361/201424259}
  {\path{doi:10.1051/0004-6361/201424259}}.

\bibitem{eckert17b}
D.~{Eckert}, M.~{Gaspari}, M.~S. {Owers}, E.~{Roediger}, S.~{Molendi},
  F.~{Gastaldello}, S.~{Paltani}, S.~{Ettori}, et~al.
\newblock {Deep Chandra observations of the stripped galaxy group falling into
  Abell 2142}.
\newblock {\em \aap}, 605:A25, September 2017.
\newblock \href {http://arxiv.org/abs/1705.05844} {\path{arXiv:1705.05844}},
  \href {http://dx.doi.org/10.1051/0004-6361/201730555}
  {\path{doi:10.1051/0004-6361/201730555}}.

\bibitem{DeGrandi16}
S.~{De Grandi}, D.~{Eckert}, S.~{Molendi}, M.~{Girardi}, E.~{Roediger},
  M.~{Gaspari}, F.~{Gastaldello}, S.~{Ghizzardi}, et~al.
\newblock {A textbook example of ram-pressure stripping in the Hydra A/A780
  cluster}.
\newblock {\em \aap}, 592:A154, August 2016.
\newblock \href {http://arxiv.org/abs/1602.07148} {\path{arXiv:1602.07148}},
  \href {http://dx.doi.org/10.1051/0004-6361/201526641}
  {\path{doi:10.1051/0004-6361/201526641}}.

\bibitem{zhuravleva18}
I.~{Zhuravleva}, S.~W. {Allen}, A.~{Mantz}, and N.~{Werner}.
\newblock {Gas Perturbations in the Cool Cores of Galaxy Clusters: Effective
  Equation of State, Velocity Power Spectra, and Turbulent Heating}.
\newblock {\em \apj}, 865:53, September 2018.
\newblock \href {http://arxiv.org/abs/1707.02304} {\path{arXiv:1707.02304}},
  \href {http://dx.doi.org/10.3847/1538-4357/aadae3}
  {\path{doi:10.3847/1538-4357/aadae3}}.

\bibitem{urban14}
O.~{Urban}, A.~{Simionescu}, N.~{Werner}, S.~W. {Allen}, S.~{Ehlert},
  I.~{Zhuravleva}, R.~G. {Morris}, A.~C. {Fabian}, et~al.
\newblock {Azimuthally resolved X-ray spectroscopy to the edge of the Perseus
  Cluster}.
\newblock {\em \mnras}, 437:3939--3961, February 2014.
\newblock \href {http://arxiv.org/abs/1307.3592} {\path{arXiv:1307.3592}},
  \href {http://dx.doi.org/10.1093/mnras/stt2209}
  {\path{doi:10.1093/mnras/stt2209}}.

\bibitem{schuecker04}
P.~{Schuecker}, A.~{Finoguenov}, F.~{Miniati}, H.~{B{\"o}hringer}, and U.~G.
  {Briel}.
\newblock {Probing turbulence in the Coma galaxy cluster}.
\newblock {\em \aap}, 426:387--397, November 2004.
\newblock \href {http://arxiv.org/abs/astro-ph/0404132}
  {\path{arXiv:astro-ph/0404132}}, \href
  {http://dx.doi.org/10.1051/0004-6361:20041039}
  {\path{doi:10.1051/0004-6361:20041039}}.

\bibitem{eckert17}
D.~{Eckert}, M.~{Gaspari}, F.~{Vazza}, F.~{Gastaldello}, A.~{Tramacere},
  S.~{Zimmer}, S.~{Ettori}, and S.~{Paltani}.
\newblock {On the Connection between Turbulent Motions and Particle
  Acceleration in Galaxy Clusters}.
\newblock {\em \apjl}, 843:L29, July 2017.
\newblock \href {http://arxiv.org/abs/1705.02341} {\path{arXiv:1705.02341}},
  \href {http://dx.doi.org/10.3847/2041-8213/aa7c1a}
  {\path{doi:10.3847/2041-8213/aa7c1a}}.

\bibitem{Siegel2018}
S.~R. {Siegel}, J.~{Sayers}, A.~{Mahdavi}, M.~{Donahue}, J.~{Merten},
  A.~{Zitrin}, M.~{Meneghetti}, K.~{Umetsu}, et~al.
\newblock {Constraints on the Mass, Concentration, and Nonthermal Pressure
  Support of Six CLASH Clusters from a Joint Analysis of X-Ray, SZ, and Lensing
  Data}.
\newblock {\em \apj}, 861:71, July 2018.
\newblock \href {http://arxiv.org/abs/1612.05377} {\path{arXiv:1612.05377}},
  \href {http://dx.doi.org/10.3847/1538-4357/aac5f8}
  {\path{doi:10.3847/1538-4357/aac5f8}}.

\bibitem{nandra13}
K.~{Nandra}, D.~{Barret}, X.~{Barcons}, A.~{Fabian}, J.-W. {den Herder},
  L.~{Piro}, M.~{Watson}, C.~{Adami}, et~al.
\newblock {The Hot and Energetic Universe: A White Paper presenting the science
  theme motivating the Athena+ mission}.
\newblock {\em ArXiv e-prints}, June 2013.
\newblock \href {http://arxiv.org/abs/1306.2307} {\path{arXiv:1306.2307}}.

\bibitem{Meidinger14}
N.~{Meidinger}, K.~{Nandra}, M.~{Plattner}, M.~{Porro}, A.~{Rau}, A.~E.
  {Santangelo}, C.~{Tenzer}, and J.~{Wilms}.
\newblock {The wide field imager instrument for Athena}.
\newblock In {\em Space Telescopes and Instrumentation 2014: Ultraviolet to
  Gamma Ray}, volume 9144 of {\em \procspie}, page 91442J, July 2014.
\newblock \href {http://dx.doi.org/10.1117/12.2054490}
  {\path{doi:10.1117/12.2054490}}.

\bibitem{gaskin15}
J.~A. {Gaskin}, M.~C. {Weisskopf}, A.~{Vikhlinin}, H.~D. {Tananbaum}, S.~R.
  {Bandler}, M.~W. {Bautz}, D.~N. {Burrows}, A.~D. {Falcone}, et~al.
\newblock {The X-ray Surveyor Mission: a concept study}.
\newblock In {\em UV, X-Ray, and Gamma-Ray Space Instrumentation for Astronomy
  XIX}, volume 9601 of {\em \procspie}, page 96010J, August 2015.
\newblock \href {http://dx.doi.org/10.1117/12.2190837}
  {\path{doi:10.1117/12.2190837}}.

\bibitem{Sayers2019}
Jack {Sayers}, Alfredo {Monta{\~n}a}, Tony {Mroczkowski}, Grant~W. {Wilson},
  Michael {Zemcov}, Adi {Zitrin}, Nath{\'a}lia {Cibirka}, Sunil {Golwala},
  et~al.
\newblock {Imaging the Thermal and Kinematic Sunyaev-Zel'dovich Effect Signals
  in a Sample of Ten Massive Galaxy Clusters: Constraints on Internal Velocity
  Structures and Bulk Velocities}.
\newblock {\em arXiv e-prints}, page arXiv:1812.06926, Dec 2018.
\newblock \href {http://arxiv.org/abs/1812.06926} {\path{arXiv:1812.06926}}.

\bibitem{Bertoldi2018}
F.~{Bertoldi}.
\newblock {The Atacama Large Aperture Submm/mm Telescope (AtLAST) Project}.
\newblock In {\em Atacama Large-Aperture Submm/mm Telescope (AtLAST)}, page~3,
  January 2018.
\newblock \href {http://dx.doi.org/10.5281/zenodo.1158842}
  {\path{doi:10.5281/zenodo.1158842}}.

\bibitem{deBreuck2018}
Carlos De~Breuck.
\newblock Site considerations for atlast, January 2018.
\newblock URL: \url{https://doi.org/10.5281/zenodo.1158848}, \href
  {http://dx.doi.org/10.5281/zenodo.1158848}
  {\path{doi:10.5281/zenodo.1158848}}.

\bibitem{Hargrave2018}
Peter Hargrave.
\newblock Atlast telescope design working group report, January 2018.
\newblock URL: \url{https://doi.org/10.5281/zenodo.1159025}, \href
  {http://dx.doi.org/10.5281/zenodo.1159025}
  {\path{doi:10.5281/zenodo.1159025}}.

\bibitem{Klaassen2018}
P.~{Klaassen} and J.~{Geach}.
\newblock {Galactic Science Case for AtLAST}.
\newblock In {\em Atacama Large-Aperture Submm/mm Telescope (AtLAST)}, page~20,
  January 2018.
\newblock \href {http://dx.doi.org/10.5281/zenodo.1159041}
  {\path{doi:10.5281/zenodo.1159041}}.

\bibitem{Mroczkowski2018}
T.~{Mroczkowski} and O.~{Noroozian}.
\newblock {AtLAST Instrumentation Considerations and Overview}.
\newblock In {\em Atacama Large-Aperture Submm/mm Telescope (AtLAST)}, page~26,
  January 2018.
\newblock \href {http://dx.doi.org/10.5281/zenodo.1159053}
  {\path{doi:10.5281/zenodo.1159053}}.

\bibitem{Otarola2019}
A.~{Otarola}, C.~{De Breuck}, T.~{Travouillon}, S.~{Matsushita}, L.-{\AA}.
  {Nyman}, A.~{Wootten}, S.~J.~E. {Radford}, M.~{Sarazin}, et~al.
\newblock {Precipitable Water Vapor, Temperature, and Wind Statistics At Sites
  Suitable for mm and Submm Wavelength Astronomy in Northern Chile}.
\newblock {\em \pasp}, 131(4):045001, April 2019.
\newblock \href {http://arxiv.org/abs/1902.04013} {\path{arXiv:1902.04013}},
  \href {http://dx.doi.org/10.1088/1538-3873/aafb78}
  {\path{doi:10.1088/1538-3873/aafb78}}.

\bibitem{mroczkowski19b}
T.~{Mroczkowski}, D.~{Nagai}, P.~{Andreani}, M.~{Arnaud}, J.~{Bartlett},
  N.~{Battaglia}, K.~{Basu}, E.~{Bulbul}, et~al.
\newblock {A High-resolution SZ View of the Warm-Hot Universe}.
\newblock {\em arXiv e-prints}, March 2019.
\newblock \href {http://arxiv.org/abs/1903.02595} {\path{arXiv:1903.02595}}.

\bibitem{nagai13}
D.~{Nagai}, E.~T. {Lau}, C.~{Avestruz}, K.~{Nelson}, and D.~H. {Rudd}.
\newblock {Predicting Merger-induced Gas Motions in {$\Lambda$}CDM Galaxy
  Clusters}.
\newblock {\em \apj}, 777:137, November 2013.
\newblock \href {http://arxiv.org/abs/1307.2251} {\path{arXiv:1307.2251}},
  \href {http://dx.doi.org/10.1088/0004-637X/777/2/137}
  {\path{doi:10.1088/0004-637X/777/2/137}}.

\bibitem{zuhone16}
J.~A. {ZuHone}, M.~{Markevitch}, and I.~{Zhuravleva}.
\newblock {Mapping the Gas Turbulence in the Coma Cluster: Predictions for
  Astro-H}.
\newblock {\em \apj}, 817:110, February 2016.
\newblock \href {http://arxiv.org/abs/1505.07848} {\path{arXiv:1505.07848}},
  \href {http://dx.doi.org/10.3847/0004-637X/817/2/110}
  {\path{doi:10.3847/0004-637X/817/2/110}}.

\bibitem{nelson18}
D.~{Nelson}, V.~{Springel}, A.~{Pillepich}, V.~{Rodriguez-Gomez}, P.~{Torrey},
  S.~{Genel}, M.~{Vogelsberger}, R.~{Pakmor}, et~al.
\newblock {The IllustrisTNG Simulations: Public Data Release}.
\newblock {\em arXiv e-prints}, December 2018.
\newblock \href {http://arxiv.org/abs/1812.05609} {\path{arXiv:1812.05609}}.

\bibitem{bulbul2012}
G.~E. {Bulbul}, R.~K. {Smith}, A.~{Foster}, J.~{Cottam}, M.~{Loewenstein},
  R.~{Mushotzky}, and R.~{Shafer}.
\newblock {High-resolution XMM-Newton Spectroscopy of the Cooling Flow Cluster
  A3112}.
\newblock {\em \apj}, 747:32, March 2012.
\newblock \href {http://arxiv.org/abs/1110.4422} {\path{arXiv:1110.4422}},
  \href {http://dx.doi.org/10.1088/0004-637X/747/1/32}
  {\path{doi:10.1088/0004-637X/747/1/32}}.

\bibitem{Sanders.ea:13a}
J.~S. {Sanders} and A.~C. {Fabian}.
\newblock {Velocity width measurements of the coolest X-ray emitting material
  in the cores of clusters, groups and elliptical galaxies}.
\newblock {\em \mnras}, 429:2727--2738, March 2013.
\newblock \href {http://arxiv.org/abs/1212.1259} {\path{arXiv:1212.1259}},
  \href {http://dx.doi.org/10.1093/mnras/sts543}
  {\path{doi:10.1093/mnras/sts543}}.

\bibitem{Pinto.ea:15}
C.~{Pinto}, J.~S. {Sanders}, N.~{Werner}, J.~{de Plaa}, A.~C. {Fabian}, Y.-Y.
  {Zhang}, J.~S. {Kaastra}, A.~{Finoguenov}, et~al.
\newblock {Chemical Enrichment RGS cluster Sample (CHEERS): Constraints on
  turbulence}.
\newblock {\em \aap}, 575:A38, March 2015.
\newblock \href {http://arxiv.org/abs/1501.01069} {\path{arXiv:1501.01069}},
  \href {http://dx.doi.org/10.1051/0004-6361/201425278}
  {\path{doi:10.1051/0004-6361/201425278}}.

\bibitem{Bambic.ea:18}
C.~J. {Bambic}, C.~{Pinto}, A.~C. {Fabian}, J.~{Sanders}, and C.~S. {Reynolds}.
\newblock {Limits on turbulent propagation of energy in cool-core clusters of
  galaxies}.
\newblock {\em \mnras}, 478:L44--L48, July 2018.
\newblock \href {http://arxiv.org/abs/1803.08175} {\path{arXiv:1803.08175}},
  \href {http://dx.doi.org/10.1093/mnrasl/sly060}
  {\path{doi:10.1093/mnrasl/sly060}}.

\bibitem{hitomi16}
{Hitomi Collaboration}, F.~{Aharonian}, H.~{Akamatsu}, F.~{Akimoto}, S.~W.
  {Allen}, N.~{Anabuki}, L.~{Angelini}, K.~{Arnaud}, et~al.
\newblock {The quiescent intracluster medium in the core of the Perseus
  cluster}.
\newblock {\em \nat}, 535:117--121, July 2016.
\newblock \href {http://arxiv.org/abs/1607.04487} {\path{arXiv:1607.04487}},
  \href {http://dx.doi.org/10.1038/nature18627}
  {\path{doi:10.1038/nature18627}}.

\bibitem{Hitomi18_turbulence}
{Hitomi Collaboration}, F.~{Aharonian}, H.~{Akamatsu}, F.~{Akimoto}, S.~W.
  {Allen}, L.~{Angelini}, M.~{Audard}, H.~{Awaki}, et~al.
\newblock {Atmospheric gas dynamics in the Perseus cluster observed with
  Hitomi}.
\newblock {\em \pasj}, 70:9, March 2018.
\newblock \href {http://arxiv.org/abs/1711.00240} {\path{arXiv:1711.00240}},
  \href {http://dx.doi.org/10.1093/pasj/psx138}
  {\path{doi:10.1093/pasj/psx138}}.

\bibitem{ZuHone2018}
J.~A. {ZuHone}, E.~D. {Miller}, E.~{Bulbul}, and I.~{Zhuravleva}.
\newblock {What Do the Hitomi Observations Tell Us About the Turbulent
  Velocities in the Perseus Cluster? Probing the Velocity Field with Mock
  Observations}.
\newblock {\em \apj}, 853:180, February 2018.
\newblock \href {http://arxiv.org/abs/1708.07206} {\path{arXiv:1708.07206}},
  \href {http://dx.doi.org/10.3847/1538-4357/aaa4b3}
  {\path{doi:10.3847/1538-4357/aaa4b3}}.

\bibitem{takahashi14}
T.~{Takahashi}, K.~{Mitsuda}, R.~{Kelley}, F.~{Aharonian}, H.~{Akamatsu},
  F.~{Akimoto}, S.~{Allen}, N.~{Anabuki}, et~al.
\newblock {The ASTRO-H X-ray astronomy satellite}.
\newblock In {\em Space Telescopes and Instrumentation 2014: Ultraviolet to
  Gamma Ray}, volume 9144 of {\em \procspie}, page 914425, July 2014.
\newblock \href {http://arxiv.org/abs/1412.1356} {\path{arXiv:1412.1356}},
  \href {http://dx.doi.org/10.1117/12.2055681} {\path{doi:10.1117/12.2055681}}.

\bibitem{hitomi18}
{Hitomi Collaboration}, F.~{Aharonian}, H.~{Akamatsu}, F.~{Akimoto}, S.~W.
  {Allen}, L.~{Angelini}, M.~{Audard}, H.~{Awaki}, et~al.
\newblock {Atmospheric gas dynamics in the Perseus cluster observed with
  Hitomi}.
\newblock {\em \pasj}, 70:9, March 2018.
\newblock \href {http://arxiv.org/abs/1711.00240} {\path{arXiv:1711.00240}},
  \href {http://dx.doi.org/10.1093/pasj/psx138}
  {\path{doi:10.1093/pasj/psx138}}.

\bibitem{lau17}
E.~T. {Lau}, M.~{Gaspari}, D.~{Nagai}, and P.~{Coppi}.
\newblock {Physical Origins of Gas Motions in Galaxy Cluster Cores:
  Interpreting Hitomi Observations of the Perseus Cluster}.
\newblock {\em \apj}, 849:54, November 2017.
\newblock \href {http://arxiv.org/abs/1705.06280} {\path{arXiv:1705.06280}},
  \href {http://dx.doi.org/10.3847/1538-4357/aa8c00}
  {\path{doi:10.3847/1538-4357/aa8c00}}.

\bibitem{bourne17}
M.~A. {Bourne} and D.~{Sijacki}.
\newblock {AGN jet feedback on a moving mesh: cocoon inflation, gas flows and
  turbulence}.
\newblock {\em \mnras}, 472:4707--4735, December 2017.
\newblock \href {http://arxiv.org/abs/1705.07900} {\path{arXiv:1705.07900}},
  \href {http://dx.doi.org/10.1093/mnras/stx2269}
  {\path{doi:10.1093/mnras/stx2269}}.

\bibitem{tashiro18}
M.~{Tashiro}, H.~{Maejima}, K.~{Toda}, R.~{Kelley}, L.~{Reichenthal},
  J.~{Lobell}, R.~{Petre}, M.~{Guainazzi}, et~al.
\newblock {Concept of the X-ray Astronomy Recovery Mission}.
\newblock In {\em Society of Photo-Optical Instrumentation Engineers (SPIE)
  Conference Series}, volume 10699 of {\em Society of Photo-Optical
  Instrumentation Engineers (SPIE) Conference Series}, page 1069922, July 2018.
\newblock \href {http://dx.doi.org/10.1117/12.2309455}
  {\path{doi:10.1117/12.2309455}}.

\bibitem{Guainazzi2018}
M.~{Guainazzi} and M.~S. {Tashiro}.
\newblock {The Hot Universe with XRISM and Athena}.
\newblock {\em arXiv e-prints}, July 2018.
\newblock \href {http://arxiv.org/abs/1807.06903} {\path{arXiv:1807.06903}}.

\bibitem{ota18}
N.~{Ota}, D.~{Nagai}, and E.~T. {Lau}.
\newblock {Constraining hydrostatic mass bias of galaxy clusters with
  high-resolution X-ray spectroscopy}.
\newblock {\em \pasj}, 70:51, June 2018.
\newblock \href {http://arxiv.org/abs/1507.02730} {\path{arXiv:1507.02730}},
  \href {http://dx.doi.org/10.1093/pasj/psy040}
  {\path{doi:10.1093/pasj/psy040}}.

\bibitem{simionescu19}
A.~{Simionescu}, J.~{ZuHone}, I.~{Zhuravleva}, E.~{Churazov}, M.~{Gaspari},
  D.~{Nagai}, N.~{Werner}, E.~{Roediger}, et~al.
\newblock {Constraining Gas Motions in the Intra-Cluster Medium}.
\newblock {\em arXiv e-prints}, January 2019.
\newblock \href {http://arxiv.org/abs/1902.00024} {\path{arXiv:1902.00024}}.

\bibitem{barret16}
D.~{Barret}, T.~{Lam Trong}, J.-W. {den Herder}, L.~{Piro}, X.~{Barcons},
  J.~{Huovelin}, R.~{Kelley}, J.~M. {Mas-Hesse}, et~al.
\newblock {The Athena X-ray Integral Field Unit (X-IFU)}.
\newblock In {\em Space Telescopes and Instrumentation 2016: Ultraviolet to
  Gamma Ray}, volume 9905 of {\em \procspie}, page 99052F, July 2016.
\newblock \href {http://arxiv.org/abs/1608.08105} {\path{arXiv:1608.08105}},
  \href {http://dx.doi.org/10.1117/12.2232432} {\path{doi:10.1117/12.2232432}}.

\bibitem{barret18}
D.~{Barret}, T.~{Lam Trong}, J.-W. {den Herder}, L.~{Piro}, M.~{Cappi},
  J.~{Houvelin}, R.~{Kelley}, J.~M. {Mas-Hesse}, et~al.
\newblock {The ATHENA X-ray Integral Field Unit (X-IFU)}.
\newblock In {\em Society of Photo-Optical Instrumentation Engineers (SPIE)
  Conference Series}, volume 10699 of {\em Society of Photo-Optical
  Instrumentation Engineers (SPIE) Conference Series}, page 106991G, July 2018.
\newblock \href {http://arxiv.org/abs/1807.06092} {\path{arXiv:1807.06092}},
  \href {http://dx.doi.org/10.1117/12.2312409} {\path{doi:10.1117/12.2312409}}.

\bibitem{ettori13}
S.~{Ettori}, G.~W. {Pratt}, J.~{de Plaa}, D.~{Eckert}, J.~{Nevalainen}, E.~S.
  {Battistelli}, S.~{Borgani}, J.~H. {Croston}, et~al.
\newblock {The Hot and Energetic Universe: The astrophysics of galaxy groups
  and clusters}.
\newblock {\em ArXiv e-prints}, June 2013.
\newblock \href {http://arxiv.org/abs/1306.2322} {\path{arXiv:1306.2322}}.

\bibitem{roncarelli18}
M.~{Roncarelli}, M.~{Gaspari}, S.~{Ettori}, V.~{Biffi}, F.~{Brighenti},
  E.~{Bulbul}, N.~{Clerc}, E.~{Cucchetti}, et~al.
\newblock {Measuring turbulence and gas motions in galaxy clusters via
  synthetic Athena X-IFU observations}.
\newblock {\em \aap}, 618:A39, October 2018.
\newblock \href {http://arxiv.org/abs/1805.02577} {\path{arXiv:1805.02577}},
  \href {http://dx.doi.org/10.1051/0004-6361/201833371}
  {\path{doi:10.1051/0004-6361/201833371}}.

\bibitem{cucchetti18}
E.~{Cucchetti}, E.~{Pointecouteau}, P.~{Peille}, N.~{Clerc}, E.~{Rasia},
  V.~{Biffi}, S.~{Borgani}, L.~{Tornatore}, et~al.
\newblock {Athena X-IFU synthetic observations of galaxy clusters to probe the
  chemical enrichment of the Universe}.
\newblock {\em \aap}, 620:A173, December 2018.
\newblock \href {http://arxiv.org/abs/1809.08903} {\path{arXiv:1809.08903}},
  \href {http://dx.doi.org/10.1051/0004-6361/201833927}
  {\path{doi:10.1051/0004-6361/201833927}}.

\end{thebibliography}
